\RequirePackage{snapshot}
\documentclass[conference]{IEEEtran}

\usepackage[utf8]{inputenc}
\usepackage[T1]{fontenc}
\usepackage[english]{babel}

\usepackage{xcolor}
\usepackage{graphicx}
\usepackage{xspace}
\usepackage{cite}
\usepackage{booktabs}
\usepackage{enumitem}

\usepackage{cleveref} %

\def\Snospace~{\S{}}

\newcommand*{\cf}[1][]  {cf.\@\xspace\ifx\\#1\\\else\Cref{#1}\fi}

\makeatletter
\newcommand*{\etc}{%
    \@ifnextchar{.}%
        {etc}%
        {etc.\@\xspace}%
}
\makeatother

\makeatletter
\newcommand*{\etal}{%
    \@ifnextchar{.}%
        {et al}%
        {et al.\@\xspace}%
}
\makeatother

\title{SecureCloud: Secure Big Data Processing in Untrusted Clouds}

\usepackage{tikz}
\usepackage{hyperref}
\newcommand\copyrighttext{  \footnotesize \textcopyright 2017 IEEE.
    Personal use of this material is permitted.
    Permission from IEEE must be obtained for all other uses,
    in any current or future media, including reprinting/republishing this
    material for advertising or promotional purposes, creating new collective
    works, for resale or redistribution to servers or
    lists, or reuse of any copyrighted component of this work in other works.
    Pre-print version. Presented in the \href{https://www.date-conference.com/date17/}{Design, Automation \& Test in Europe Conference \& Exhibition (DATE '17), Lausanne - Switzerland, 2017}. For the final published paper, refer to the DOI: \href{https://doi.org/10.23919/DATE.2017.7926999}{10.23919/DATE.2017.7926999}}
\newcommand\copyrightnotice{\begin{tikzpicture}[remember picture,overlay]
\node[anchor=south,yshift=10pt,fill=yellow!20] at (current page.south) {\fbox{\parbox{\dimexpr\textwidth-\fboxsep-\fboxrule\relax}{\copyrighttext}}};
\end{tikzpicture}}
\begin{document}

\author{\IEEEauthorblockN{
Florian Kelbert\IEEEauthorrefmark{1},
Franz Gregor\IEEEauthorrefmark{2},
Rafael Pires\IEEEauthorrefmark{3},
Stefan Köpsell\IEEEauthorrefmark{2},
Marcelo Pasin\IEEEauthorrefmark{3},\\
Aurélien Havet\IEEEauthorrefmark{3},
Valerio Schiavoni\IEEEauthorrefmark{3},
Pascal Felber\IEEEauthorrefmark{3},
Christof Fetzer\IEEEauthorrefmark{2},
Peter Pietzuch\IEEEauthorrefmark{1}}\medskip
\IEEEauthorblockA{\IEEEauthorrefmark{1}Imperial College London, United Kingdom,
\texttt{\{fkelbert, prp\}@imperial.ac.uk}}
\IEEEauthorblockA{\IEEEauthorrefmark{2}TU Dresden, Germany,
\texttt{\{firstname.lastname\}@tu-dresden.de}}
\IEEEauthorblockA{\IEEEauthorrefmark{3}University of Neuchâtel, Switzerland,
\texttt{\{firstname.lastname\}@unine.ch}} }

\maketitle
\copyrightnotice

\begin{abstract}
  We present the SecureCloud EU Horizon 2020 project, whose goal is to enable
  new big data applications that use sensitive data in the cloud without
  compromising data security and privacy. For this, SecureCloud designs and
  develops a layered architecture that allows for (i)~the secure creation and
  deployment of secure micro-services; (ii)~the secure integration of
  individual micro-services to full-fledged big data applications; and
  (iii)~the secure execution of these applications within untrusted cloud
  environments. To provide security guarantees, SecureCloud leverages novel
  security mechanisms present in recent commodity CPUs, in particular,
  Intel's Software Guard Extensions~(SGX). SecureCloud applies this
  architecture to big data applications in the context of smart grids. We
  describe the SecureCloud approach, initial results, and considered use
  cases.
\end{abstract}

\section{Introduction}
\label{sec:intro}

Despite a steady increase in cloud adoption over the past few years, some
challenges remain. Confidentiality, integrity, and availability of
applications and their data are of immediate concern to 
organisations that use cloud computing. This is particularly true for
organisations that must comply with strict policies, including those which
process personal data or that support society's most critical infrastructures,
such as finance, health care, and smart grids. The goal of 
SecureCloud is to address such concerns by providing solutions
that allow for the secure processing of sensitive data within untrusted clouds.

The primary area of application of the developed solutions 
is in the field of critical infrastructures, whose operators
have legitimate concerns about the dependability of applications hosted in
third-party clouds. Despite security guarantees given by cloud operators,
dependability concerns increasingly become a barrier to the broad adoption of 
cloud computing. 
The cloud therefore becomes itself a critical
infrastructure for which we need to provide sufficient guarantees so
that we can justifiably place our trust in their hosted applications.

The overall goal of our work is to develop a platform that enables the 
dependable
implementation, deployment and execution of critical applications within
untrusted cloud environments. Our objectives are:
\begin{enumerate}[wide,nosep]
\item substantially improve the state-of-the-art in cloud dependability 
  by developing innovative and effective mechanisms to
  enforce security, covering integrity and confidentiality, as well as
  availability and reliability;

\item seamlessly integrate new dependability features into a standard cloud
  stack to encourage easy migration of critical (as well as non-critical)
  applications to the cloud without compromising application dependability; and

\item convincingly validate and demonstrate the benefits of our approach by
  applying it to realistic and demanding big data use cases in the domain of
  critical infrastructures (smart grids).

\end{enumerate}

\section{Existing Approaches to Cloud Security}
\label{sec:related}

A modern public cloud is home to a hardware and software stack consisting of
many devices, a large codebase and large frameworks that are often
immature, rapidly evolving, and full of bugs and configuration errors that can
be exploited by attackers. The challenge for operators is to convince potential
clients that it is safe to execute their applications and store their data in
such a dangerous environment.

One approach taken by operators is to define a Trusted Computing Base (TCB)
within their stack~\cite{tpm}. Typically, the TCB includes most of the basic
middleware, operating system~(OS), and networking facilities of the data
centre, as well as its hardware platform.  Establishing the credibility of the
TCB amounts to verifying the correctness and security of a large and complex
hardware and software system. High costs aside and
repeating it on a continuous basis as the hardware and software evolve, the
goal of a truly ``trustworthy'' TCB has proven
elusive~\cite{McCune:2008:LYG:1353534.1346285,Sadeghi:2006:TIN:1179474.1179487}.
Even if
it were possible to remove all bugs from the TCB, this alone would not ensure
their security---given insufficient physical security or a malicious
system administrator, there could still be
unauthorized access to customer data when it is unencrypted in memory.

An approach focused specifically on securing application data from access by
both external and internal malicious agents is based on homomorphic
encryption---a technique intended to allow encrypted computations to be carried
out on encrypted data~\cite{fhe,Tebaa2012}. Since unencrypted computations and 
data would never be present in the cloud, they would never be exposed to attacks.  
As of now, the realisation of homomorphic encryption is
proving as elusive~\cite{naehrig2011can} and impractical for virtually
all real-world applications due to its immense overheads, precluding
its use in timely demanding applications.

A third approach is the use of specialised security co-processors~\cite{ibm_secure_cpu}. 
Such processors consider the chip area as a trust boundary, treating everything outside as subject to attacks and potentially compromised. 
The instructions and data are stored encrypted in the 
memory. 
Once read by the processor, they are decrypted and the 
instructions carried out on plain-text data. 
Since everything outside the chip can be tampered
with, the processor never outputs plaintext data, encrypting it before writing 
to the system bus. 
While secure processors provide good security guarantees,
specialised hardware use is counter to the general
principle of data centre ``scale out'' notion, which advocates the use of large
numbers of commodity components.

\section{SecureCloud Approach}
\label{sec:approach}

\subsection{Intel SGX and Small Trusted Computing Base}

The innovative approach to cloud dependability pursued by SecureCloud uses
novel cryptographic hardware found in upcoming commodity CPUs---in particular,
Intel SGX~\cite{costanintel,anati2013innovative}.
It allows protected execution on encrypted data where the corresponding
plaintext is only known inside the processor. 
The \emph{enclave} is a secure area in which the processing of the plaintext data happens. 
Applications are thus isolated not only from other applications but also from the underlying operating system
and hypervisor. 
Users run sensitive applications in public clouds without unconditionally trusting the cloud provider.

\subsection{Layered Architecture}
\label{sec:approach:layered}

The SecureCloud approach tackles secure processing in the cloud using SGX from a full stack perspective. 
The architecture builds on several layers and various technologies:

\smallskip

\noindent
\emph{(1)~Secure containers for QoS-aware applications.}  While many
hardware extensions for CPU/IO/memory virtualization have reduced the
overhead of virtual machines, container frameworks such as
Docker\footnote{https://www.docker.com/} are still more efficient,
although less secure since their security is 
directly linked to the underlying host OS. 
We address this tension by designing and
implementing a solution for secure containers. 
The developed components also monitor hardware usage to detect resource 
bottlenecks and allow for accounting and billing.

\smallskip

\noindent
\emph{(2)~Dependable micro-services for the cloud} utilise these secure
containers. For this, we design and implement a framework and related interfaces
which allow for the development of arbitrary, yet secure, micro-services. We further implement a few
common micro-services.

\smallskip

\noindent
\emph{(3)~Secure distributed big data applications} on the basis of
secure micro-services address big data processing.
The developed big data processing components are leveraged by the application
level demonstrators in the context of smart grids as described in
\Cref{sec:usecases}.
Examples of developed components are secure structured data stores, map/reduce
based computations, schedulers, as well as components for efficient transmission
of large amounts of data.

\section{SecureCloud Infrastructure}
\label{sec:infrastructure}

\begin{figure}[tb]
	\centering
	\includegraphics[width=0.9\linewidth]{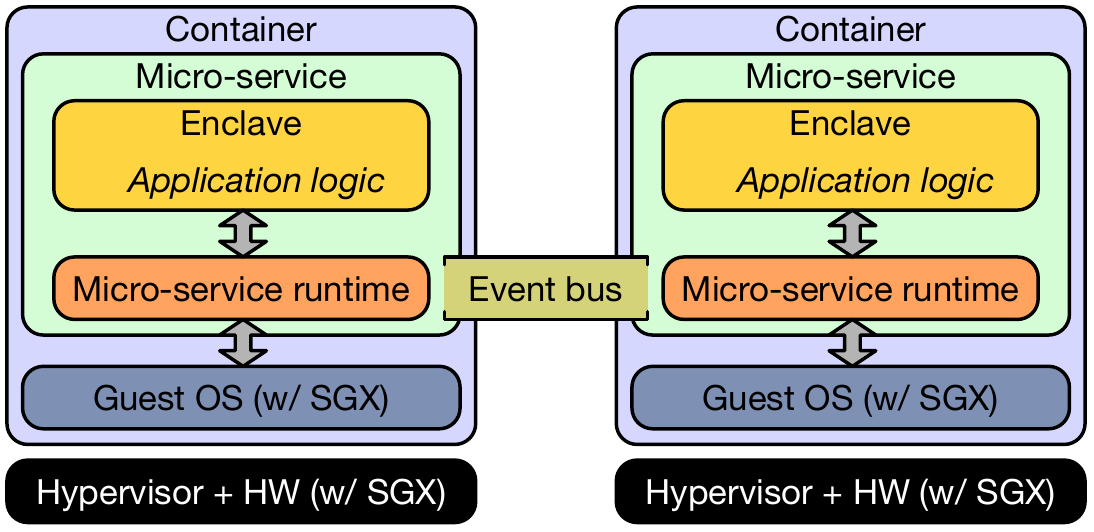}
    \vspace{-2pt}
	\caption{SecureCloud applications consist of a set of micro-services
          connected by an event bus. Our main focus is to enhance the security
          of containers. These containers may share the host with virtual
          machines, i.e., the system still contains a hypervisor.}
	\label{fig:arch-overview}
    \vspace{-6pt}
\end{figure}

\Cref{fig:arch-overview} shows the baseline infrastructure of SecureCloud.
An application consists of a set of micro-services
connected via an event bus.  The application logic of each micro-service lives
within an enclave. The micro-service runtime exists outside of the
enclave. These runtime functions only access encrypted data. 
Encryption and decryption of
this data is performed automatically and transparently within the
enclave. This approach limits the amount of code added to the TCB.

To deploy the micro-service, we offer \emph{secure containers} on top of the
untrusted stack of the cloud provider: a secure container adds confidentiality
and integrity to Docker containers.
This enables system administrators to build
secure container images within a trusted environment and to run them in an 
untrusted cloud.  To facilitate the
creation of secure containers, we designed and developed a Secure Linux 
Container Environment~(SCONE)~\cite{arnautov2016scone} that secures existing 
applications with SGX. 

To the micro-service, SCONE exposes an \emph{external system call based
interface}, which is shielded from attacks.  To protect itself from user space
attacks, SCONE performs sanity checks and copies all memory-based return values
to the inside of the enclave before passing the arguments to the micro-service.
SCONE further (i)~transparently encrypts and authenticates data that is
processed via file descriptors, and (ii)~provides acceptable performance by
implementing tailored threading and an asynchronous system call interface.

SCONE integrates with existing Docker environments, and ensures that
secure containers are compatible with standard containers. The host OS,
however, must include a Linux SGX driver and, to boost performance, a SCONE
kernel module.

With respect to Docker container deployment and scheduling, SecureCloud contributes GenPack~\cite{Havet2017}, a scheduling and monitoring framework that leverages principles from generational \emph{garbage collection} (GC)~\cite{Lieberman:1983:RGC:358141.358147}.
The core idea of GenPack is to partition the servers into several groups, named \emph{generations}.
It combines runtime monitoring of system containers to learn their requirements and properties, and a scheduler that manages different generations of servers.
\section{Preliminary Results}
\label{sec:results}

SecureCloud already developed prototypes: an unmodified
Docker ecosystem to securely deploy micro-services
(\Cref{sec:secure-docker}) and SCBR~\cite{Pires16:SCBR}, a secure
messaging system over content-based routing that allows to securely
hook-up individual micro-services to full-fledged applications (\Cref{sec:secure-routing}).

\subsection{Secure Docker Containers}
\label{sec:secure-docker}

Micro-services need a runtime environment. In line with
\Cref{sec:approach:layered}, we chose to deploy
micro-services using a containerized \emph{Docker} infrastructure, 
a currently popular and widely used platform.  
Each micro-service is executed within a \emph{secure container}---a dedicated
Docker container that runs their code protected by SGX enclaves. From
the perspective of the Docker infrastructure, secure containers are
indistinguishable from regular containers.

The integration of secure containers with Docker requires changes to the image
build process. We further provide a wrapper for the Docker client,
called \emph{SCONE client}, which provides functionalities for spawning secure
containers and for secure communication with containers.
Note that we do not require modifications to the Docker Engine or its API. 
SCONE supports the typical Docker workflow: developers publish a Docker image
featuring their micro-service; end-users can customize this image by
adding additional file system layers.

\begin{figure}[t]
\centering
\includegraphics[width=1\columnwidth]{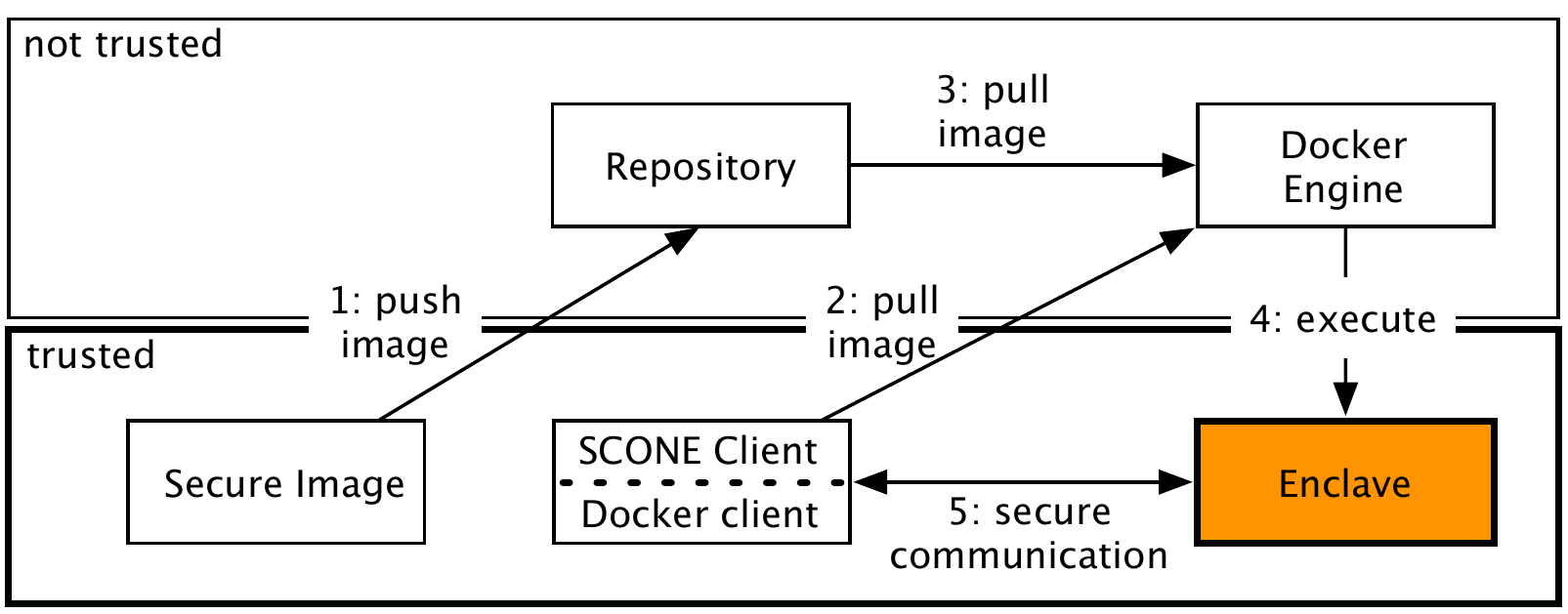}
\vspace{-10pt}
\caption{Using secure containers with Docker}
\label{fig:docker}
\vspace{-6pt}
\end{figure}

We assume that Docker images contain micro-services that are created in a
trusted environment (see \Cref{fig:docker}). The image creator must be familiar
with the security-relevant aspects of the micro-service, e.g., which files must
be protected. Next we explain the secure container image creation process.

First, the image creator builds a protected executable of the
micro-service by statically compiling against
its library dependencies and the SCONE library---a C library that
ensures that the application is only executed inside of an SGX enclave. SCONE
does not support shared libraries by design to ensure that all enclave code is
verified by SGX upon enclave creation.

Second, the image creator uses the SCONE client to protect the image's file
system~(FS). The SCONE client encrypts all files that must be protected and
creates an FS protection file, which contains the message authentication
codes~(MACs) for file chunks as well as the encryption keys. The FS protection
file itself is then encrypted and added to the image.

Lastly, the secure image is published using the standard Docker registry. As all
security-relevant parts of the image are protected by the FS protection file, we
do not need to trust the Docker registry. To allow for the secure image's
further customization, the image creator would only sign the FS
protection file, but not encrypt it. This way, the image's integrity
is ensured. Confidentiality can then only be assured after finishing the
customization process.

Each secure container requires a startup configuration file~(SCF).  The SCF
contains keys to encrypt standard I/O streams, the hash and encryption key of
the FS protection file, application arguments, as well as environment
variables. Only an enclave whose identity has been verified can access the SCF,
which is received through a TLS-protected connection that is established during
enclave startup. 

\smallskip

\subsection{Secure Content Based Routing}
\label{sec:secure-routing}

Content-based routing~(CBR) is a flexible and powerful paradigm for scalable
communication among distributed processes. It decouples data producers from
consumers, and routes messages based on their content. 
Although extensively studied~\cite{PS:Survey:03}, the publish/subscribe communication model still fails to reach wide deployment due to privacy concerns.

To perform efficient routing, a CBR router must see the content of the messages,
as well as subscriptions by data consumers, a clear threat to
privacy. 
We provide a secure CBR engine called \emph{SCBR}~\cite{Pires16:SCBR}.
It exploits SGX to perform this matching step. Hence, the
compute-intensive CBR operations can operate on decrypted data shielded by
enclaves and leverage efficient matching algorithms.

Outside of secure enclaves, both publications and subscriptions are encrypted
and signed, thus protecting the system from unauthorised parties observing or
tampering with the information. SCBR combines a key exchange protocol and a
state-of-the-art routing engine to provide both security and performance while
executing under the protection of an enclave. 
Performance is enhanced by storing subscriptions in data structures that
exploit containment relations between filters. Therefore, a reduced
number of comparisons is required whenever a message must be
matched against them.

We evaluate SCRB with several workloads to
observe the sources of performance overheads and trade-offs of SGX.
Our time measurements %
inside/outside of enclaves highlighted performance degrades when 
cache misses rate increase (i.e. data must be evicted or fetched to/from system
memory, causing SGX to perform encryption, decryption, integrity and
freshness checks).
While cache misses imposes some limited overhead, they are less critical than
memory swapping.

\begin{figure}[t]
\centering
\includegraphics[width=0.9\linewidth]{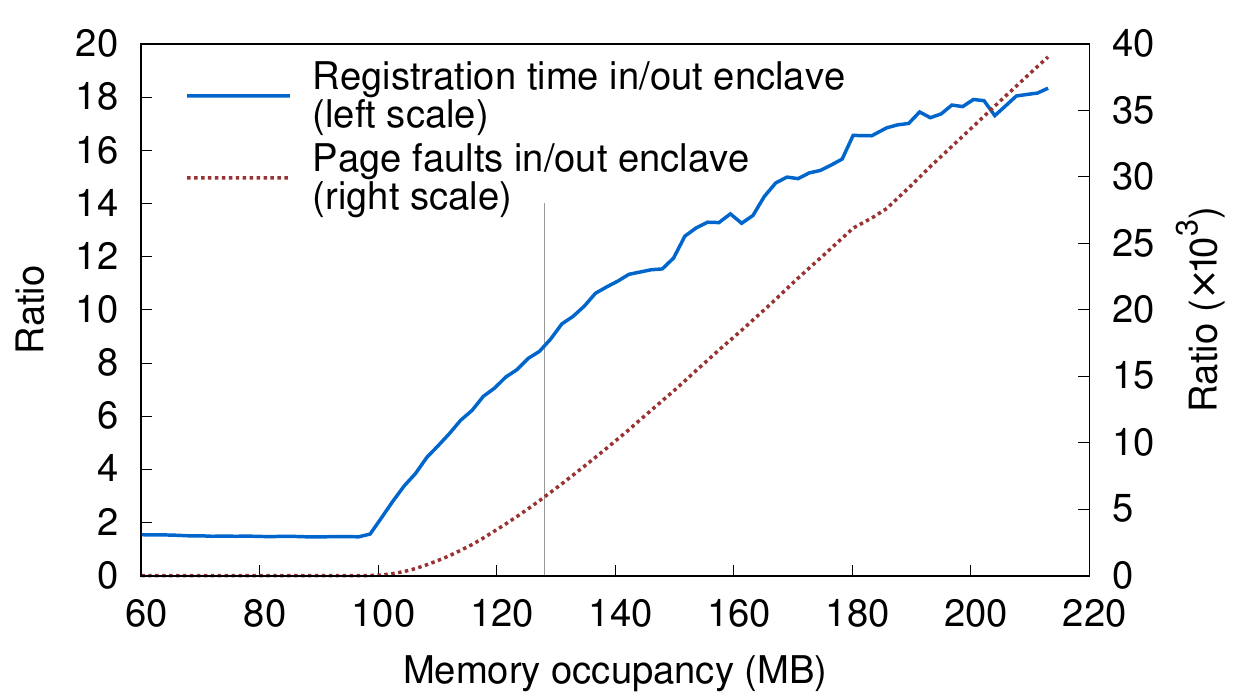}
\vspace{-2pt}
\caption{Effect of memory swapping}
\label{fig:epc}
\vspace{-6pt}
\end{figure}

Since the enclave page cache~(EPC) memory is limited to 128MB, pages must be
evicted from the protected area to the main (untrusted) memory whenever more 
space is required. 
Memory swapping is serviced by the operating system, which causes higher overheads when compared to cache misses.
\Cref{fig:epc} shows this effect by showing the combined results of matching
times when executing the same code inside and outside secure enclaves.  
Performance degrades to nearly 18$\times$ for a subscription database of
200MB. 
Even if EPC size was set to 128MB (marked by the vertical line), the
performance drop is evident before due to the use of protected memory for
SGX internal data structures.

These first results open the way for further research to minimise
memory footprint and build an enclave-efficient system.  We intend to optimise
our data structures to avoid paging and cache misses. We expect these
optimisations to further decrease the overhead of running inside an enclave.

\section{Application Use Cases}
\label{sec:usecases}

To demonstrate the need and the feasibility of using secure clouds,
the project considers use cases in the area of smart grids, which
offer opportunities to tackle many of the requirements
that sensitive big data applications may face when executing in the
cloud. 
First, they account for a growing volume of data, as 
meters and sensors for monitoring distribution and transmission grids  
continuously collect and transmit data.
Second, the energy distributors' data analyses
require access to consumers' detailed information about energy
consumption, which represents an enormous privacy risk
(their activities and behaviours can be inferred~\cite{greveler2012multimedia}).
Finally, data analysis can trigger reactions that 
interfere with the physical world (load control or consumer notifications). 
Adversaries could thus have devastating effects on the power system.

In our first use case, smart meters collect detailed power consumption data
from residential and industrial consumers. 
Collecting data at sub-minute granularities enables for sophisticated applications, such as
power theft prevention and early detection of power quality issues. 
Nowadays, such applications are deployed on dedicated servers maintained by utilities and system integrators. 
Hence, several customers cannot use them, because it would require a large data storage and processing infrastructure. 
Cloud computing offers such an infrastructure. 
Nevertheless, once this data is under control of a cloud provider, an adversary who compromises this provider's infrastructure could gain access to them, hence the need to be stored and processed securely.

The second use case considers applications that affect energy delivery
and fault detection. For these applications, data sources may be public, but
the data needs to be reliable and the processing tasks that trigger actions in
the smart grid must be executed in a timely fashion. These applications will
be supervised using monitoring services. Orchestration services detect
anomalies within milliseconds, which requires adaptations to the
virtual infrastructure that hosts the application. This fine-granular and
highly responsive orchestration system will enforce quality-of-service
guarantees without interfering with security and privacy requirements,
and can even provide better energy efficiency.
Our experiments with GenPack~\cite{Havet2017} show that up to 23\% energy 
savings are possible for typical data-center workloads. 
\section{Conclusions}

The EU SecureCloud project designs and develops technologies to enhance the
dependability of future cloud environments that host critical infrastructures
such as the smart grid. Building upon Intel's SGX technology, the developed
solutions allow for the secure creation, deployment, and execution of big data
applications in untrusted clouds. The developed solutions are applied to  
the area of smart grids, which accounts for an ever-growing
volume of data and demands for reliable, secure, and timely data
processing. Our initial SGX-based prototypes for secure and efficient data
processing and content routing demonstrate the promise of the SecureCloud
approach.

\medskip

\noindent \emph{Acknowledgements.}  The SecureCloud project has received
funding from the European Union's Horizon 2020 research and innovation
programme and was supported by the Swiss State Secretariat for Education,
Research and Innovation~(SERI) under grant agreement number 690111.  Rafael
Pires is also sponsored by CNPq, National Counsel of Technological and
Scientific Development, Brazil.

\bibliographystyle{IEEEtran}

\end{document}